\documentclass[proceedings]{JHEP3}

\PrHEP{PrHEP hep2001}                   
\conference{International Europhysics Conference on HEP}

\def\beq{\begin{equation}}   \def\eeq{\end{equation}}
\def\bea{\begin{eqnarray}}  \def\eea{\end{eqnarray}} \def\nn{\nonumber}
\def\noi{\noindent}

\title{Remarks on sum rules in the heavy quark limit of QCD}

\author{A. Le Yaouanc, \speaker{L. Oliver}, O. P\`ene and J.-C. Raynal\\
Laboratoire de Physique Th\'eorique (UMR CNRS 8627),
Universit\'e de Paris-Sud, b\^atiment 210, 91405 Orsay Cedex, France
}

\author{V. Mor\'enas\\ Laboratoire de Physique Corpusculaire,
Universit\'e Blaise Pascal, CNRS/IN2P3, 63000 Aubi\`ere Cedex,
France}

\abstract{We underline a problem existing in the heavy quark limit of
QCD concerning the rates of semileptonic $B$ decays into $P$-wave
$D_J(j)$ mesons, where $j = {1 \over 2}$ (wide states) or $j = {3 \over
2}$ (narrow states). The leading order sum rules of Bjorken and
Uraltsev suggest $\Gamma \left [ \bar{B} \to D_{0,1} \left ( {1 \over
2} \right ) \ell \nu \right ] \ll \Gamma \left [ \bar{B} \to D_{1,2}
\left ( {3 \over 2} \right ) \ell \nu \right ]$, in contradiction with
experiment. The same trend follows also from a sum rule for the
subleading $1/m_Q$ curent matrix element correction $\xi_3(1)$. The
problem is made explicit in relativistic quarks models \`a la Bakamjian
and Thomas, that give a transparent physical interpretation
of the latter as due, not to a ${\bf L\cdot S}$ force, but to the Wigner
rotation of the light quark spin. We point out moreover that the
selection rule for decay constants of $j = {3 \over 2}$ states,
$f_{3/2} = 0$, predicts, assuming the model of factorization, the
opposite hierarchy $\Gamma \left [ \bar{B} \to \bar{D}_{s_{1,2}} \left ( {3 \over
2} \right ) D^{(*)}\right ] \ll \Gamma \left [ \bar{B} \to \bar{D}_{s_{0,1}}
\left ( {1 \over 2} \right ) D^{(*)} \right ]$.}

\begin{document}
It has been recently pointed out by Uraltsev \cite{1r} that the
function $T(\varepsilon, {\bf v}, {\bf v} - {\bf v}')$, Fourier
transform of $<B^*({\bf v} - {\bf v}')|T(J^+(0)J(x))|B^*(0)>$ (where
the momentum transfer is $- m_Q {\bf v}$), can be decomposed into
symmetric and antisymmetric parts in ${\bf v}$, ${\bf v}'$,
$h_{\pm}(\varepsilon )$. The zero order moment of $h_+(\varepsilon )$
leads to Bjorken sum rule \cite{2r}~:
\beq \label{1e} \rho^2 = {1 \over 4} + \sum\nolimits_n \left |
\tau^{(n)}_{1/2}(1) \right |^2 + 2 \sum\nolimits_n \left | \tau^{(n)}_{3/2}(1)
\right |^2 \eeq
\noi while the zero order moment of $h_-(\varepsilon )$ gives a new sum
rule~:
\beq \label{2e} \sum\nolimits_n \left | \tau^{(n)}_{3/2}(1) \right |^2 - \sum\nolimits_n
\left | \tau^{(n)}_{1/2}(1) \right |^2  = {1 \over 4} \quad . \eeq
\noi Note that this relation is to be contrasted with the
non-relativistic quark model with spin-orbit independent interactions
\cite{3r}~: $\tau^{(n)}_{3/2}(1) = \tau^{(n)}_{1/2}(1)$. Indeed,
Uraltsev SR corresponds to subleading corrections in the
non-relativistic expansion. In these relations $\tau_j(w)$ are the two independent IW functions for
the $B \to D_J(j)$ transitions ($J = 0,1$, for $j = {1 \over 2}$ and $J =
1,2$ for $j = {3 \over 2}$). From both sum rules it follows the lower
bound for the elastic IW function $\rho^2 \geq {3 \over 4}$. This bound was obtained some years ago within Bakamjian-Thomas
quark models (covariant for $m_Q \to \infty$ and satisfying IW scaling)
\cite{4r}. Combining both SR one obtains $\sum_n | \tau^{(n)}_{1/2}(1) |^2 = {1
\over 3} \left ( \rho^2 - {3 \over 4} \right )$. This simple relation shows that, since $\rho^2$ is of $O(1)$,
there is little room left for the sum $\sum_n  |
\tau^{(n)}_{1/2}(1) |^2$, compared to $\sum_n  |
\tau^{(n)}_{3/2}(1)  |^2$. The same qualitative trend is suggested from a SR obtained for the
subleading function quantity $\xi_3(1)$ \cite{5r}~:
\beq \label{5e} \xi_3(1) = 2 \sum\nolimits_n \Delta E_{3/2}^{(n)} \left |
\tau^{(n)}_{3/2}(1) \right |^2 - 2 \sum\nolimits_n \Delta E_{1/2}^{(n)} \left |
\tau^{(n)}_{1/2}(1) \right |^2 \eeq
\noi $\xi_3(1)$ and $\bar{\Lambda} = m_B - m_b$ describe, at zero recoil, the
subleading corrections to the ground state current matrix elements due
to the perturbations of the current, the quantities denoted by $L_i(1)$
$(i = 4,5,6)$ in the notation of Falk and Neubert \cite{6r}. Note that $\xi_3(1)$ is identical to $\bar{\Sigma}$, defined in \cite{1r}. On the other hand, $\bar{\Lambda}$ satisfies Voloshin SR \cite{7r}
\beq \label{6e} {\bar{\Lambda} \over 2} =  \sum\nolimits_n \Delta E_{1/2}^{(n)}
\left | \tau^{(n)}_{3/2}(1) \right |^2 + 2 \sum\nolimits_n \Delta E_{3/2}^{(n)}
\left | \tau^{(n)}_{3/2}(1) \right |^2 \quad . \eeq
\noi If one combines both subleading SR together with the QCDSR
result (ignoring short distance corrections) \cite{8r} $\xi_3(1) = {\bar{\Lambda} \over 3}$ one obtains the ratio 
$\sum_n  \Delta E_{3/2}^{(n)}  |
\tau^{(n)}_{3/2}(1)  |^2/\sum_n \Delta E_{1/2}^{(n)}
 | \tau^{(n)}_{1/2}(1)  |^2$ $= 4$. Since in quarkonia the ${\bf L \cdot S}$ force turns out to be small,
the level spacings $\Delta E_j^{(n)}$ are roughly independent of $j$.
One concludes again that one expects the hierarchy $\sum_n | \tau^{(n)}_{3/2}(1)  |^2$  $\gg
\sum_n  | \tau^{(n)}_{1/2}(1)  |^2$. Making the natural hypothesis that this hierarchy holds for the $n
= 0$ states, one expects the inequalities \cite{9r}
\bea \label{10e} &&\Gamma \left [ \bar{B}_d \to D_{1,2} \left ( {3 \over
2} \right ) \ell \nu \right ] \gg \Gamma \left [ \bar{B}_d \to D_{0,1}
\left ( {1 \over 2} \right ) \ell \nu \right ] \nn \\ &&\Gamma \left [
\bar{B}_d \to D_{1,2} \left ( {3 \over 2} \right ) \pi \right ] \gg
\Gamma \left [ \bar{B}_d \to D_{0,1} \left ( {1 \over 2} \right ) \pi
\right ] \eea
\noi where the second inequality follows assuming factorization for $B
\to D_J(j) \pi$ $(j = {1 \over 2}, {3 \over 2})$, a natural hypothesis
taking into account the recent work in QCD on factorization in two-body
non-leptonic decays when a light meson is emitted \cite{10r}. However, experiment does not show this trend, as shown by recent data
(see \cite{11r} in Table 1 and \cite{12r}). Up to now, our considerations have been model-independent. \par

To illustrate different aspects, let us consider the Bakamjian-Thomas
quark models \cite{4r}. These are relativistic quark models with an
interacting fixed number of constituents, the states forming
representations of the Poincar\'e Group. It has been shown that, in the
heavy quark limit, current form factors are covariant, and Isgur-Wise
scaling is satisfied \cite{4r}. Moreover, Bjorken sum rule holds
\cite{13rnew}. This is a class of relativistic models, each of them
characterized by a mass operator $M$ Ansatz, that describes the
dynamics at rest. One can thus make use of the different spectroscopic
models proposed in the literature and compute the elastic IW function
$\xi( w)$ and the transition $P$-wave IW functions $\tau_{1/2}(w)$,
$\tau_{3/2}(w)$ as has been done in ref. \cite{13r}. The calculation of decay constants of
heavy mesons for which HQET scaling is also satisfied has been also
performed within the same approach \cite{14r}. It has recently been shown that Uraltsev SR holds as well in the BT
framework \cite{9r}, that also gives a transparent physical
interpretation of the difference $\tau^{(n)}_{3/2}(1) -
\tau^{(n)}_{1/2}(1)$. In these classes of models one has, in terms of
wave functions at rest \cite{13r}
\beq \label{11e} \tau^{(n)}_{3/2}(1) - \tau^{(n)}_{1/2}(1) \cong {1
\over (2 \pi )^2 \sqrt{3}} \int p^2 dp \left [ p \varphi_{L=1}^{(n)}
(p)\right ]^* \ {p \over p_0 + m} \ \varphi (p) \eeq
\noi where, assuming small ${\bf L \cdot S}$ coupling,
$\varphi^{(n)}_{1/2}(p) \cong \varphi^{(n)}_{3/2}(p) =
\varphi_{L=1}^{(n)} (p)$, and $m$, $p$ and $p_0$ are
the light quark mass, momentum and energy in the hadron rest frame. In
this limit, the difference is due to the relativistic structure of the
current matrix elements in terms of wave functions, namely due to the
light spectator {\it quark Wigner rotations}, i.e. a relativistic
kinematical effect from the center-of-mass boost. The difference is
large~: for a constituent quark mass $m \cong 0.3$~GeV, ${p \over p_0 +
m}$ is of $O(1)$. It must be emphasized that this effect is independent
of a possible difference that could come from the small spin-orbit
force that yields a difference between the internal wave functions
at rest $\varphi^{(n)}_{1/2}(p)$, $\varphi^{(n)}_{3/2}(p)$. Uraltsev SR follows in the BT scheme as follows. A change
of variables is performed between the quark momenta $({\bf p}_1, {\bf
p}_2) \to ({\bf P}, {\bf k}_2)$, ${\bf P}$ being the center-of-mass
momentum, and ${\bf k}_2$ the internal relative momentum. A current
matrix element writes
\bea 
\label{12e}
&&<{\bf v}' |J_{\mu}(0)|{\bf v}> = \sum\nolimits_{s'_1s_1} \bar{u}_{s'_1} \ \gamma_{\mu} \ u_{s_1} \int d{\bf p}_2 {\sqrt{(p_i \cdot v) (p'_i \cdot v')} \over p_2^0} \nn \\
&&\sum\nolimits_{s'_2s_2} \varphi '^{*}_{s'_2s_2}({\bf k}'_2) \left [ D\left ( R'^{-1}_2 R_2 \right ) \right ]_{s'_2s_2} \ \varphi_{s_1s_2} ({\bf k}_2) 
\eea
\noi where $u_{s'_1} \gamma_{\mu} u_{s_1}$ expresses the
fact that quark 1 is the active heavy quark. The first term under the
integral is the Jacobian of the change of variables. The functions
$\varphi$ and $\varphi '$ are wave functions at rest, dependent on
relative momenta and Pauli spinors. The matrix $D\left (
R_2^{'-1}R_2\right )$ is the Wigner rotation on the spectator quark
spin due to the boosts on initial and final states. Considering, following Uraltsev \cite{1r}, the hadronic tensor 
\beq \label{13e} h_{00}^{+10}\left ({\bf v}_f, {\bf v}', {\bf v}_i\right ) = \sum\nolimits_n
<B^{*(+1)} ({\bf v}_f) |V_0(0)|n({\bf v}')> \ <n({\bf v}') |V_0(0)|
B^{*(0)} ({\bf v}_i)> \ . \eeq

An expansion in the BT scheme yields
\beq \label{14e} < n({\bf v}') |V_0(0)| O({\bf v})> \cong (n|0) + {1
\over 2} ({\bf v}' - {\bf v}) \cdot \left ( n\left | -i \left ( p_2^0 {\bf
r}_2 + {\bf r}_2 p_2^0 \right ) + {i\left (\sigma_2 \times {\bf
p}_2\right ) \over p_2^0 + m} \right ] 0\right ) \ .\eeq
\noi Where the operator $i\left ( p_2^0 {\bf r}_2 + {\bf r}_2
p_2^0\right )$ becomes the dipole operator $2im{\bf r}_2$ in the
non-relativistic limit and ${i\left (\sigma_2 \times {\bf p}_2\right )
\over p_2^0 + m}$ is a Wigner rotation. For $V_0$, the active quark $\bar{u}_{s'_1}\gamma_0 u_{s_1}$
cannot produce the necessary spin flip $B^{*(0)} \to B^{*(+1)}$, but
this is produced by the Wigner rotation of the spectator light quark.
One obtains
\beq \label{16e} h_{00}^{+10}\left ({\bf v}_f, {\bf v}', {\bf
v}_i\right ) \cong {1 \over 4} v_f^z {1 \over \sqrt{2}} \left ( v'^x -
i v'^y \right ) \eeq
\noi expressing quark-hadron duality. The same hadronic tensor in terms
of phenomenological IW functions $\tau_j(w)$ yields
\beq \label{17e} h_{00}^{+10}\left ( {\bf v}_f, {\bf v}', {\bf
v}_i\right ) \cong v_f^z {1 \over \sqrt{2}} \left ( v'^x - i v'^y
\right ) \left [ \sum\nolimits_n \left | \tau^{(n)}_{3/2}(1) \right |^2 - \sum\nolimits_n
\left | \tau^{(n)}_{1/2}(1) \right |^2 \right ]\eeq
\noi and hence Uraltsev SR follows. Tables 1 and 2 show calculations in the BT scheme for different
dynamics at rest provided by different studies of meson spectroscopy.
One can see that there is a quick convergence of Bjorken and
Uraltsev sum rules, almost saturated by $n = 0$ states.

\TABLE{\begin{tabular}{|c|c|c|c|}
\hline
Quark-antiquark &Godfrey, Isgur \protect{\cite{15r}}  &Cea, Colangelo, &Isgur, Scora \\
Potential &$(Q\bar{Q},Q\bar{q},q\bar{q})$ &Cosmai, Nardulli \protect{\cite{16r}} &Grisntein, Wise \protect{\cite{17r}}    \\
\hline
$\left | \tau^{(0)}_{1/2}(1) \right |^2$ &0.051 &0.004 &0.117 \\
 \hline
$\left | \tau^{(0)}_{3/2}(1) \right |^2$ &0.291 &0.265 &0.305 \\
\hline
${\scriptstyle {1 \over 4} + \left | \tau^{(0)}_{1/2}(1) \right |^2 + 2 \left | \tau^{(0)}_{3/2}(1) \right |^2}$ &0.882 &0.790 &1.068 \\
\hline
$\rho^2$ &1.023 &0.98 &1.283 \\
\hline
${\scriptstyle \left | \tau^{(0)}_{3/2}(1) \right |^2 - \left | \tau^{(0)}_{1/2}(1) \right |^2}$ &0.240 &0.261 &0.233 \\
\hline
\end{tabular}
\caption{Contributions of the $n=0$ states to the Sum Rules in BT models}
} 

\TABLE{\begin{tabular}{|c|c|c|c|c|}
\hline
Semileptonic Mode &Godfrey-Isgur   &Cea et al. &Isgur et al. &Expt. \protect{\cite{11r}}\\
  \hline
$B \to D\ell \nu$ &2.36 \% &2.45 \% &1.94 \% &$(2.1 \pm 0.2) \%$ \\
 \hline
 $B \to D^*\ell \nu$ &6.86 \% &7.02 \% &6.07 \% &$(5.3 \pm 0.8) \%$ \\
 \hline
 & & &  &(a) $(2.4 \pm 1.1) \times 10^{-3}$ \\
$B \to D_2\ell \nu$ &$7.0 \times 10^{-3}$ &$6.5 \times 10^{-3}$ &$7.7 \times 10^{-3}$ &(b) $(4.4 \pm 2.4) \times 10^{-3}$ \\ 
& & & &(c) $(3.0 \pm 3.4) \times 10^{-3}$ \\
\hline
& & &  &(a) $(7.0 \pm 1.6) \times 10^{-3}$ \\
$B \to D_1\left ( {3 \over 2} \right ) \ell \nu$ &$4.5 \times 10^{-3}$ &$4.2 \times 10^{-3}$ &$4.9 \times 10^{-3}$ &(b) $(6.7 \pm 2.1) \times 10^{-3}$ \\ 
& & & &(c) $(5.6 \pm 1.6) \times 10^{-3}$ \\
\hline
$B \to D^*_1\left ( {1 \over 2} \right ) \ell \nu$ &$7 \times 10^{-4}$ &$4 \times 10^{-5}$ &$1.3 \times 10^{-3}$ &(d) $(2.3 \pm 0.7) \times 10^{-2}$ \\ 
\cline{1-4}
 $B \to D^*_0\left ( {1 \over 2} \right ) \ell \nu$ &$6 \times 10^{-4}$ &$4 \times 10^{-5}$ &$1.1 \times 10^{-3}$ &$\left [ D^*_0 \left ( {1 \over 2} \right ) + D_1^* \left ( {1 \over 2} \right ) \right ]$\\ 
\hline
 \end{tabular}
\caption{Comparison between semileptonic decay rates in BT models and data : (a) ALEPH, (b) DELPHI, (c) CLEO \protect{\cite{11r}} and (d) CLEO \protect{\cite{12r}}.}
}

  Let us now comment on the decay constants of $P$-wave mesons in
the heavy quark limit of QCD. In this limit, the following selection
rule was found \cite{18r} $f_{3/2}^{(n)} = 0$. Assuming the {\it model} of factorization one finds the hierarchy
in the {\it emission} of $P$-wave mesons
\beq \label{19e} \Gamma \left ( \bar{B}_d \to \bar{D}_{s_{1,2}} \left (
{3 \over 2} \right ) D^{(*)} \right ) \ll \Gamma \left ( \bar{B}_d \to
\bar{D}_{s_{0,1}} \left ( {1 \over 2} \right ) D^{(*)} \right )  \quad
. \eeq
\noi The BaBar experiment is looking for $\bar{B}_d \to
(\bar{D}K)D^{(*)}$ that could allow to test this inequality. This is
interesting, because it goes on exactly the opposite direction of the
hierarchy in the production of these mesons, discussed above, that
follows on HQET general grounds. BT quark models satisfy heavy quark scaling for the decay constants,
namely $\sqrt{M}\  f = Const.$ and the relation $f_B = f_{B^*}$,
together with the selection rule $f_{3/2} = 0$. For the allowed decay
constants $f_{1/2}$  one finds for the $n = 0$
states, within the Godfrey-Isgur spectroscopic model \cite{14r}~:
$\sqrt{M}$ $f_{1/2}^{(0)} = 0.64$~GeV$^{3/2}$, very close to the value
found for the ground state $S$-wave $\sqrt{M}$ $f^{(0)} =
0.67$~GeV$^{3/2}$ ($M$ is the mass of the heavy meson). Therefore,
assuming factorization, one predicts relations of the kind $\Gamma \left ( \bar{B}_d \to D^-_{s_{0}} \left (
{1 \over 2} \right ) D^{+} \right ) / \Gamma \left ( \bar{B}_d \to
D_{s}^- D^+  \right ) \cong 1$. 

 In conclusion, we have underlined that there is at present a
problem between the prediction in heavy quark limit of QCD (\ref{10e}) and experiment. The understanding of these rates is important, as
they are directly related to the slope of the elastic IW function
$\rho^2$. On the other hand, we have shown that Uraltsev SR is
satisfied by Bakamjian-Thomas relativistic quark models, that
provide a physical interpretation of the sum rule as due to
the light spectator {\it quark Wigner rotations}, and not to a possible
${\bf L \cdot S}$ force. Uraltsev Sum Rule yields a {\it rationale} for the
bound $\rho^2 \geq {3 \over 4}$ found within the BT quark models some
years ago. On the other hand, the heavy quark limit selection rule
$f_{3/2}^{(n)} = 0$, plus the factorization hypothesis predicts an
opposite hierarchy for the emission of $P$-wave mesons.

\end{document}